\documentclass[referee]{raa}
\usepackage{graphicx,times}
\usepackage{subfigure}
\usepackage{amssymb}
\usepackage{threeparttable}
\usepackage{epstopdf}
\usepackage{caption}
\usepackage{hyperref}
\usepackage{booktabs}
\usepackage{multirow}
\usepackage{array}
\makeatletter
\newcommand{\rmnum}[1]{\romannumeral #1}
\newcommand{\Rmnum}[1]{\expandafter\@slowromancap\romannumeral #1@}
\makeatother
\newcommand{\tabincell}[2]{\begin{tabular}{@{}#1@{}}#2\end{tabular}}

\begin{document}

    \title{Extremely Metal-Poor Star Candidates in the SDSS}

    \volnopage{Vol.0 (200x) No.0, 000--000}
    \setcounter{page}{1}

    \author{S. Y. Xu
       \inst{1,2}
    \and H. W. Zhang
        \inst{1}
    \and X. W. Liu
        \inst{1,2}}
    \institute{Department of Astronomy, School of Physics, Peking University, Beijing 100871, P.R.China; \\
         \and
              Kavli Institute of Astronomy and Astrophysics, Peking University, Beijing 100871, P.R.China {\it syxu@pku.edu.cn}}
    \date{}

\abstract{ For a sample of metal-poor stars $(-3.3\leq $ [Fe/H] $ \leq-2.2)$ that have high-resolution spectroscopic abundance determinations, we have measured equivalent widths (EW) of the Ca\,\textsc{\rmnum{2}}\,K, Mg\,\textsc{\rmnum{1}}\,b and near-infrared (NIR) Ca\,\textsc{\rmnum{2}}\,triplet lines using low-resolution spectra of the Sloan Digital Sky Survey (SDSS), calculated effective temperatures from (\emph{g} $-$ \emph{z})$_0$ color, deduced stellar surface gravities by fitting stellar isochrones, and determined metallicities based on the aforementioned quantities. Metallicities thus derived from the Ca\,\textsc{\rmnum{2}}\,K line are in much better agreement with the results determined from high-resolution spectra than the values given in the SDSS Data Release 7 (DR7). The metallicities derived from the Mg\,\textsc{\rmnum{1}}\,b lines have a large dispersion owing to the large measurement errors, whereas those deduced from the Ca\,\textsc{\rmnum{2}}\,triplet lines are too high due to both non-local thermodynamical equilibrium (NLTE) effects and measurement errors. Abundances after corrected for the NLTE effect for the Mg\,\textsc{\rmnum{1}}\,b lines and Ca\,\textsc{\rmnum{2}}\,triplet lines are also presented. Following this method, we have identified six candidates of ultra-metal-poor stars with [Fe/H] $\sim -4.0$ from a sample of 166 metal-poor star candidates. One of them, SDSS J102915+172927, was recently confirmed to be an ultra-metal-poor ([Fe/H] $< -4.0$) star with the lowest metallicity ever measured. Follow-up high-resolution spectroscopy for the other five ultra-metal-poor stars in our sample will therefore be of great interest.
}
   \authorrunning{S. Y. Xu, H. W. Zhang \& X. W. Liu}

    \maketitle

\section{Introduction}
\label{sect:intro}

Elemental abundance patterns in the cosmos result from the nuclear processes that happened in the Big Bang, stars, and the interstellar medium
(\cite{Page97}).
Measurements and analyses of stellar element abundances can help to address the formation and evolution of stars and galaxies. Searching for extremely metal-poor
(EMP, [Fe/H] $< -3$; Beers \& Christlieb~\cite{Bee05})
stars and follow-up detailed analyses of their elemental abundances can give us a deep insight of the formation of the first generation of stars and the early chemical evolution of galaxies
(Frebel~\cite{Freb10}).
To obtain accurate elemental abundance of EMP stars, one needs time-consuming high-resolution spectroscopic observations. As such, candidates of EMP stars are first selected from low to intermediate-resolution spectra for follow-up high-resolution spectroscopic observations
(Beers \& Christlieb~\cite{Bee05}).
The most iron-poor stars discovered so far are all candidates selected by low-resolution spectroscopic surveys, such as the HK survey
(\cite{Bee92})
and the Hamburg/ESO survey
(\cite{Wiso96}),
and their ultra-low metallicities are later confirmed by follow-up high-resolution spectroscopic observations, e.g., HE 0107-5240
([Fe/H] $= -5.2$; \cite{Chri02}), HE 1327-2326
([Fe/H] $= -5.4$; \cite{Freb05}; \cite{Aoki06}), and HE 0557-4840
([Fe/H] $= -4.75$; \cite{Norr07}).
Thus, reliable metallicity determinations for metal-poor stars from low to intermediate-resolution spectra is the key to find and identify more EMP stars.\\
Using spectra of a resolving power R $\sim 7500$ from the Radial Velocity Experiment (RAVE) survey,
Fulbright et al.~(\cite{Fulb10})
develop an efficient technique of identifying candidates of ultra-metal-poor stars ([Fe/H] $< -4$) from a pool of metal-poor ([Fe/H] $< -2$) stars, without the need of further observations of higher spectral resolutions. With this method, three stars with [Fe/H] $\lesssim-4$ were identified in a sample of approximately 200,000 stars.\\
The large number of low-resolution spectra provided by the SDSS and to be provided by the forthcoming LAMOST (The Large Sky Area Multi-Object Fiber Spectroscopic Telescope, also named The Guoshoujing Telescope;
\cite{Wang96}; \cite{Su98}; \cite{Xing98}; \cite{Zhao00}; \cite{Cui04}; \cite{Zhu06}; \cite{Zhao12}; c.f. \it http://www.lamost.org/website/en/)
\rm Galactic surveys
(\cite{Deng12}; Liu et al. 2012, in preparation)
motivate the need for better metallicity determinations from those spectra, in particular at low metallicity, such that reliable stellar candidates of EMP stars can be identified for follow-up high spectral resolution, high signal-to-noise ratio ($S/N$) spectroscopy. Preliminary observations of LAMOST obtained during the early commissioning phase have shown its ability of searching for metal-poor stars in the Milky Way
(\cite{Li10}; \cite{Wu10}; \cite{Ren12}).
As one of the hitherto most successful and productive surveys, the SDSS has accumulated hundreds of thousands low resolution stellar spectra that can be utilized to search for candidates of EMP stars.
The stellar atmosphere parameters released in DR7 of the SDSS
(\cite{Abaz09})
are provided by the SEGUE Stellar Parameter Pipeline (SSPP). For the DR7 version of the SSPP,  the metallicity determinations at the extremum of [Fe/H] $< -3.0$ were not well constrained due to lack of corresponding calibrators
(Smolinski et al.~\cite{Smol11}).
There is evidence that the SSPP-derived metallicities for stars below [Fe/H] $= -2.7$ can be $0.3$\,dex higher than the actual values determined by high-resolution spectroscopy
(\cite{Yan09}).
The performance of the SSPP has been significantly improved in DR8 of the SDSS
(\cite{Aiha11}),
yet the most metal deficient star found by the SSPP in this huge pool of more than half a million stars has an unexpected, rather high metallicity, [Fe/H] $=-3.9$, indicating that even the current, most up-to-date version of the SSPP might have missed the most metal-poor stars. Motivated by the success of
Fulbright et al.~(\cite{Fulb10})
in finding EMPs with the RAVE spectra, we have applied their technique to the huge data set of the SDSS spectra of even lower spectral resolution than those of RAVE. We show that the method is not only applicable to the SDSS spectra, it is even capable of finding candidates of [Fe/H] $<-4$. \\
The paper is organized as follows. In Section 2, we describe our method and apply it to a sample of 13 metal-poor stars that have both the SDSS and high-resolution spectroscopic observations. In Section 3, we extend the analysis to a large sample of low-metallicity star candidates in the SDSS data archive and search for candidates of EMP stars. Our summary follows in Section 4.

\section{Method}
\label{sect:meth}
\subsection{Selection of Sample Metal-Poor Stars with High-Resolution Spectra}
\label{sect:Sel}
The abundance data derived from high-resolution spectroscopy of known metal-poor stars can be found in the table that contains elemental abundances of $\sim\!\!1000$ metal-poor stars collected from the literature by Frebel~(\cite{Freb10}), and the Stellar Abundances for Galactic Archeology (SAGA) Database
(\cite{Sud08})
that contains 1212 metal-poor halo stars with [Fe/H] $<-2.5$. Those valuable data provide an effective way to examine our method and test the results. We select metal-poor stars with  [Fe/H] $<-2.0$ from the compilation of Frebel and the SAGA database, and identify them in the SDSS archive. In total, we find 13 metal-poor stars with the SDSS spectra. Table~\ref{Tab:HR-13}) lists the stellar parameters deduced from high-resolution spectroscopic analyses and those given in SDSS DR8. Note that there exist systematic uncertainties in the high-resolution abundance data collected from different literature studies (the original references are also listed in Table~\ref{Tab:HR-13}))
(Frebel~\cite{Freb10}).
But for the relatively small-scale systematic uncertainties (around $0.3$\,dex, Frebel~\cite{Freb10}) will not affect our conclusions, we can use this high-resolution reference sample to provide robust estimates for [Fe/H].\\
\begin{table}[h]
\centering
\begin{threeparttable}
\caption[]{Stellar Parameters of the Sample of 13 Stars. }\label{Tab:HR-13}
  \begin{tabular}{ccccp{0.1cm}cccp{0.1cm}ccc}
    \toprule
\multirow{3}{*}{\centering STAR} & \multicolumn{3}{c}{HR} & & \multicolumn{3}{c}{DR8} & & \multicolumn{3}{c}{This work} \\
  \cline{2-4} \cline{6-8} \cline{10-12}
    & \emph{T}$_{\rm eff}$ & \textrm{log}~\emph{g} & [Fe/H] & & \emph{T}$_{\rm eff}$ & \textrm{log}~\emph{g} & [Fe/H] & & \emph{T}$_{\rm eff}$ & \textrm{log}~\emph{g} & [Fe/H] \\
    & (K) & $\rm (cm\,s^{-2})$ & (dex) &  & (K) & $\rm (cm\,s^{-2})$ & (dex) && (K) & $\rm (cm\,s^{-2})$ & (dex) \\
  \midrule
SDSS2047+00\tnote{1} & $6600$ & $4.5$  & $-2.1$ & & $6316$ & $3.8$ & $-2.2$ && $6119$ & $3.8$ &$-2.6$  \\
HE2134+0001\tnote{2} & $5257$ & $3.0$  & $-2.2$ && $5436$ & $3.1$ & $-2.0$ && $5430$ & $3.4$ & $-2.2$ \\
HE1132+0125\tnote{2} & $5732$ & $3.5$ & $-2.4$ && $6186$ & $4.1$ & $-2.1$ && $6024$ & $3.8$ & $-2.4$ \\
SDSS0036-10\tnote{1} & $6500$ & $4.5$ & $-2.5$  & & $6445$ & $4.3$ & $-2.5$ && $6237$ & $4.4$ & $-2.9$ \\
SDSS0924+40\tnote{1} & $6200$ & $4.0$ & $-2.6$ && $6176$ & $4.5$ & $-2.7$ && $5973$ & $3.7$ & $-2.8$ \\
SDSS1707+58\tnote{1} & $6700$ & $4.2$ & $-2.6$ && $6581$ & $3.5$ & $-2.7$ && $6395$ & $4.4$ & $-2.9$ \\
HE1128-0823\tnote{2} & $5909$ & $3.7$  & $-2.7$ && $6083$ & $3.9$ & $-2.6$ && $6041$ & $3.7$ & $-2.6$ \\
HE0315+0000\tnote{2} & $5013$ & $2.1$ & $-2.7$ && $5031$ & $2.1$ & $-2.8$ && $4999$ & $2.0$ & $-3.0$ \\
SDSS0126+06\tnote{1} & $6600$ & $4.1$ & $-3.2$ && $6852$ & $4.7$ & $-2.8$ && $6615$ & $4.2$ & $-3.0$ \\
SDSS J234723\tnote{3} & $4600$ & $1.5$ & $-3.2$ && $5025$ & $2.4$ & $-2.5$ && $4729$ & $1.4$ & $-2.7$ \\
SDSS0817+26\tnote{1}& $6300$ & $4.0$ & $-3.2$ && $6077$ & $3.6$ & $-3.1$ && $5997$ & $3.6$ & $-3.2$ \\
SDSS1033+40\tnote{4} & $6370$ & $4.4$  & $-3.2$ && $6574$ & $4.2$ & $-3.1$ && $6309$ & $4.4$ & $-3.2$ \\
SDSS0040+16\tnote{4} & $6360$ & $4.4$ & $-3.3$ && $6665$ & $3.8$ & $-3.1$ && $6306$ & $4.4$ & $-3.3$ \\
    \bottomrule
    \end{tabular}
    \begin{tablenotes}
       \footnotesize
       \item[1] \cite{Aoki08}
       \item[2] \cite{Bark05}
       \item[3] \cite{Lai09}
       \item[4] \cite{Aoki09}
    \end{tablenotes}  
\end{threeparttable}
\end{table}
A comparison of the abundances derived from high-resolution spectra and those determined with the SSPP used in DR8, it is clear that the SSPP has overestimated the metallicities of those metal-poor stars, by $\sim\!\!0.1$\,dex on average (see also Fig.~\ref{Fig:f3d} in Section~\ref{sect:Ana}). In addition, the SSPP also seems to have overestimated the effective temperatures (by approximately $100$\,K) than those adopted in high-resolution spectroscopic analyses (c.f. Fig.~\ref{Fig:f1b} in Section~\ref{sect:Det}). The latter is probably an important factor leading to the higher metallicities of the SSPP.

\subsection{The Adopted Metal Line Features}
Given the limited $S/N$ and spectral resolution of the SDSS spectra, weak Fe\,\textsc{\rmnum{1}} lines can only be used for metallicity analysis in a limited stellar parameter range (\cite{Ren12}). The Ca\,\textsc{\rmnum{2}}\,K line at $3933.7$\,\r{A}, the Mg\,\textsc{\rmnum{1}}\,b lines at $5167.3$, $5172.7$ and $5183.6$\,\r{A}, and the Ca\,\textsc{\rmnum{2}}\,triplet lines at $8498.0$, $8542.1$ and $8662.1$\,\r{A}, are on other hand stronger and easier to measure for metal-poor stars. The Ca\,\textsc{\rmnum{2}}\,K line is probably the best indicator for the overall metallicity of a metal-poor star when only low $S/N$ and resolution spectra are available
(Frebel~\cite{Freb10}).
Generally, there exists a degeneracy amongst the stellar parameters including effective temperature, surface gravity and metallicity. A lower-metallicity, cooler giant and a higher-metallicity hotter turnoff star can have equal strengths of those lines
(Fulbright et al.~\cite{Fulb10}).
For the Ca\,\textsc{\rmnum{2}}\,triplet lines, a $100$\,K difference in temperature leads to about $0.1$\,dex difference in metallicity
(Fulbright et al.~\cite{Fulb10}),
i.e., an overestimated effective temperature will result in an overestimated metallicity. \\
For metal-poor star of [Fe/H] $\sim -1.5$, one can assume that the $\alpha$-element (Ca, Mg, etc.) abundances are enhanced by $0.4$\,dex  with respect to the Solar $\alpha$/Fe ratio
(Fulbright et al.~\cite{Fulb10}).
Although there are some outliers that have a strong enhancement of $\alpha$ element (Frebel~\cite{Freb10}, fig. 2), this assumption is generally found to be valid in most cases and is also assumed in the current work when we convert the deduced calcium or magnesium abundance to that of iron.

\subsection{Determination of Stellar Atmosphere Parameters}
\label{sect:Det}

\subsubsection{Effective Temperature}
\label{subsect:Dettef}
Being the most important factor influencing estimates of metallicities, erroneously assigned effective temperatures can result in biased elemental abundances. For our sample, the DR8 version of the SSPP produces a higher estimate of effective temperature,  $\sim 100$\,K higher than those adopted in high-resolution spectroscopic analyses.
Fulbright et al.~(\cite{Fulb10})
utilize the NIR photometry from the 2MASS survey
(\cite{Cutr03})
and optical \emph{UBV} photometry to calculate photometric effective temperature \emph{T}$_{\mathrm{phot}}$. Here we make use of the high quality SDSS \emph{ugriz} photometry and use the long-baseline (\emph{g} $-$ \emph{z} )$_0$ color as effective temperature indicator. Two calibrations are available to us.\\
Ludwig et al.~(\cite{Lud08})
obtain a theoretical relation between the (\emph{g} $-$ \emph{z})$_0$ color and the effective temperature based on model atmospheres,
\begin{equation}
\emph{T}_{\mathrm{eff}}=7126.7 - 2844.2\cdot (\emph{g} - \emph{z}\rm )_0 + 666.80\cdot (\emph{g}-\emph{z}\rm ) _0^{2} - 11.724\cdot (\emph{g} - \emph{z}\rm ) _0^{3}.
\label{eq:teff1}
\end{equation}
Smolinski et al.~(\cite{Smol11}),
based on likely cluster members and calibration, derive two color-temperature relations applicable to different metallicity regions. For the range of elemental abundance of interest here, we adopt their relation for [Fe/H] $<-1.5$,
\begin{equation}
\emph{T}_{\rm eff}=6993 - 2573\cdot (\emph{g} - \emph{z}\rm ) _0 + 530.9\cdot (\emph{g} - \emph{z}\rm ) _0^{2}.
\label{eq:teff2}
\end{equation}
The effective temperatures derived from the above two calibrations are referred to as $T_1$ and $T_2$. $T_1$ and $T_2$ are respectively $31$ and $74$\,K lower than the values adopted in high-resolution spectroscopic analyses (mostly are photometric temperatures) on average.
\begin{figure}[htbp]
   \centering \mbox{
   \subfigure[]{
   \includegraphics[width=8cm]{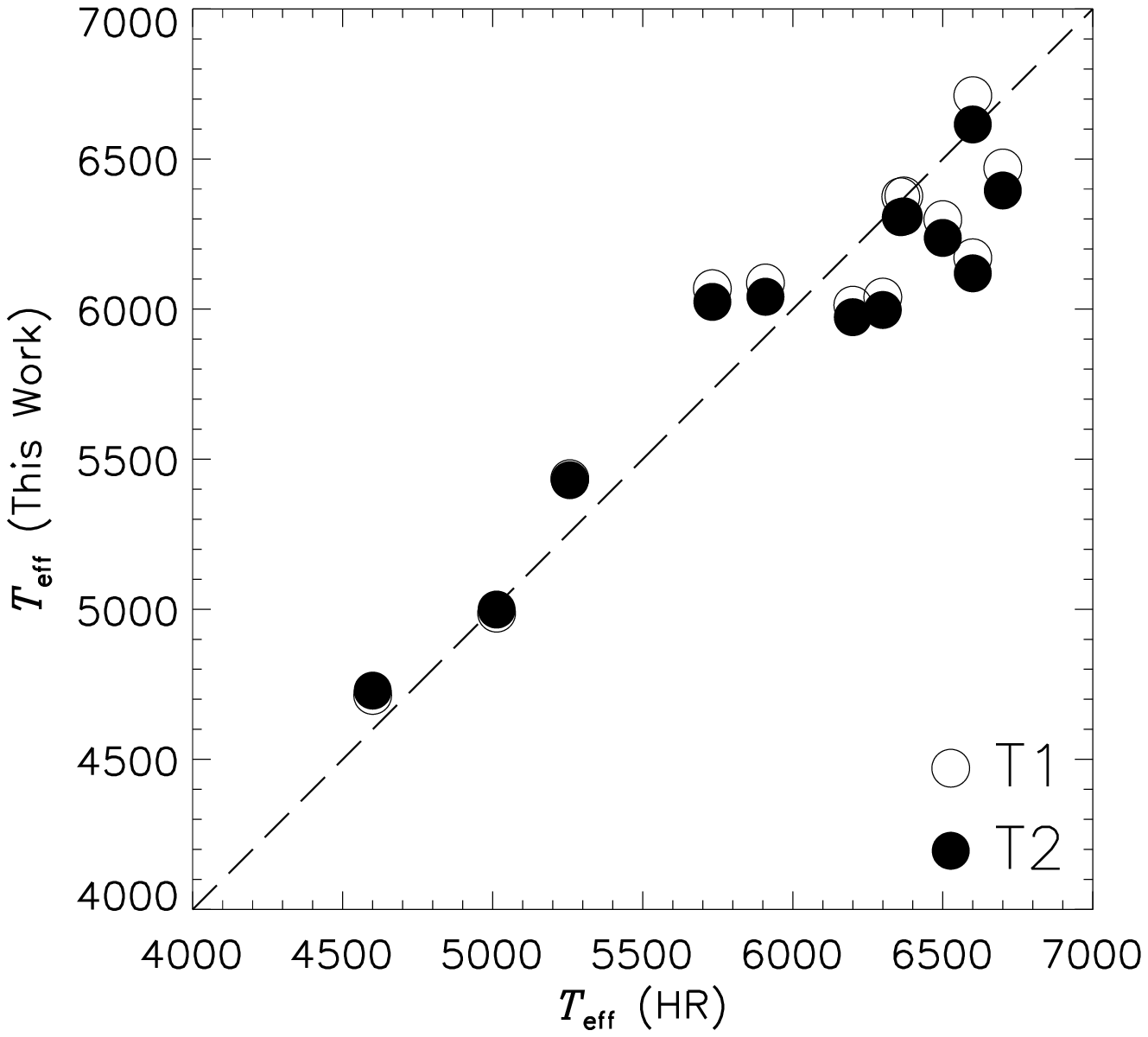}\label{Fig:f1a}}
   \quad
   \subfigure[]{
   \includegraphics[width=8cm]{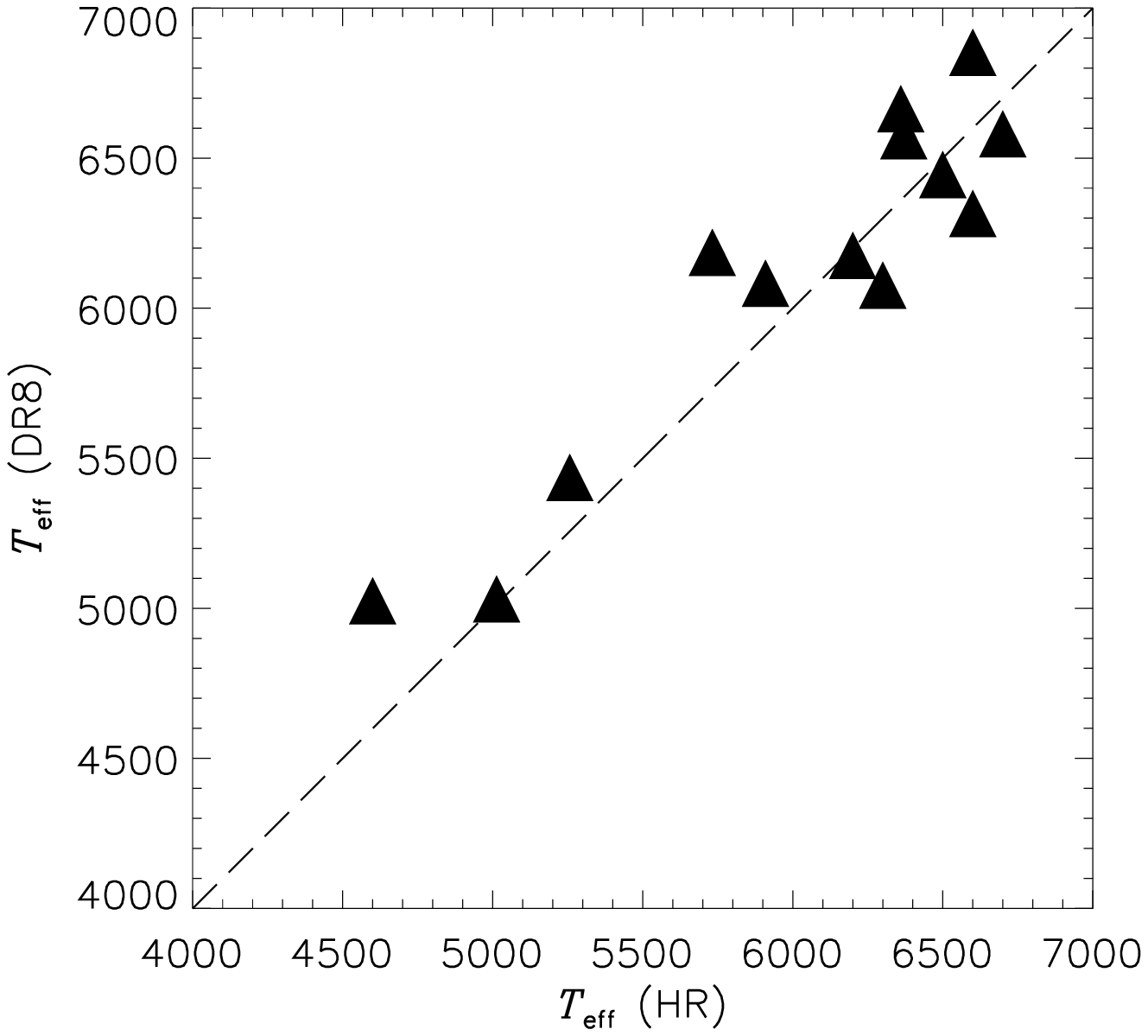}\label{Fig:f1b}}}
   \mbox{
   \subfigure[]{
   \includegraphics[width=8cm]{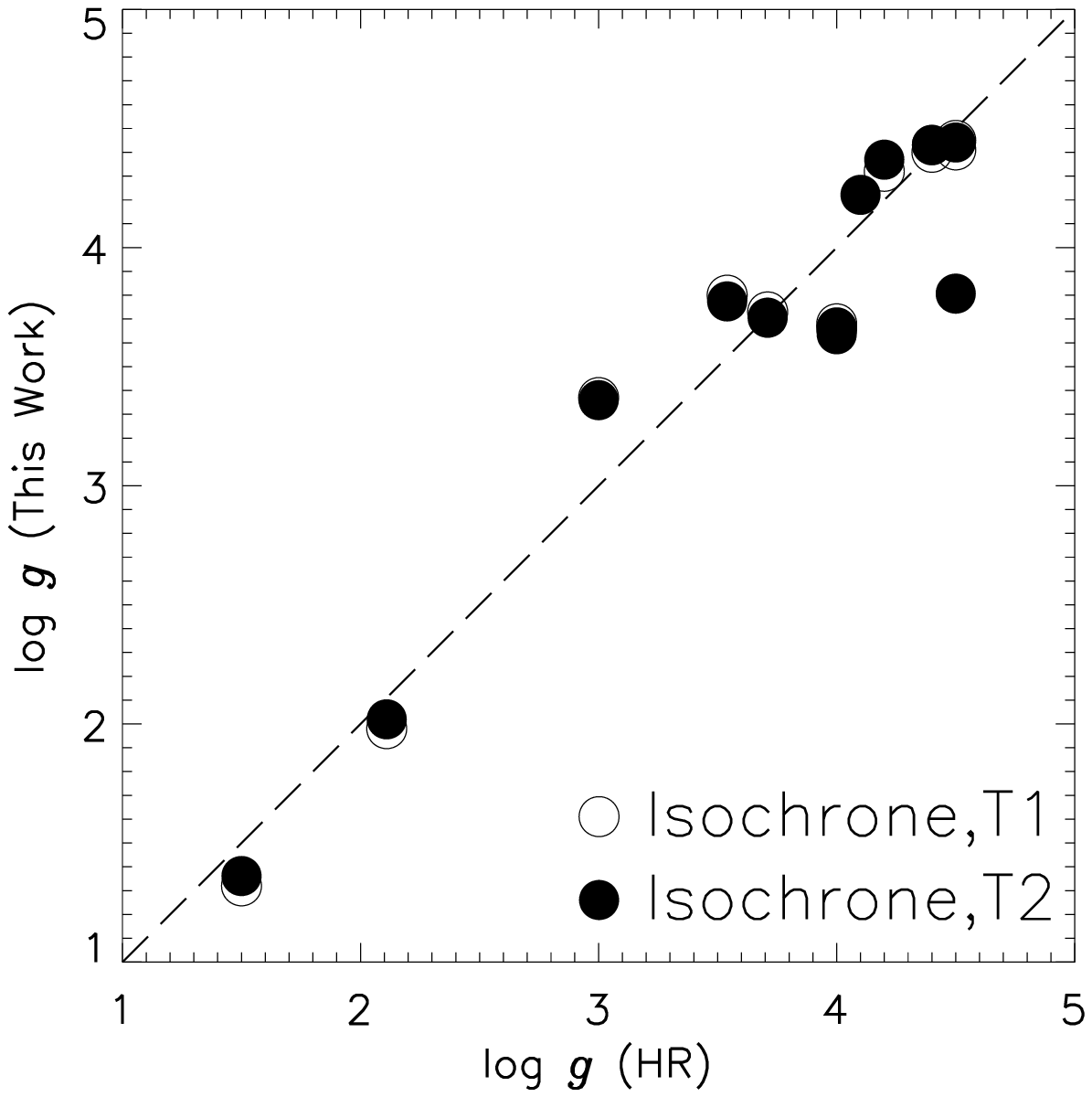}\label{Fig:f1c}}
   \quad
   \subfigure[]{
   \includegraphics[width=8cm]{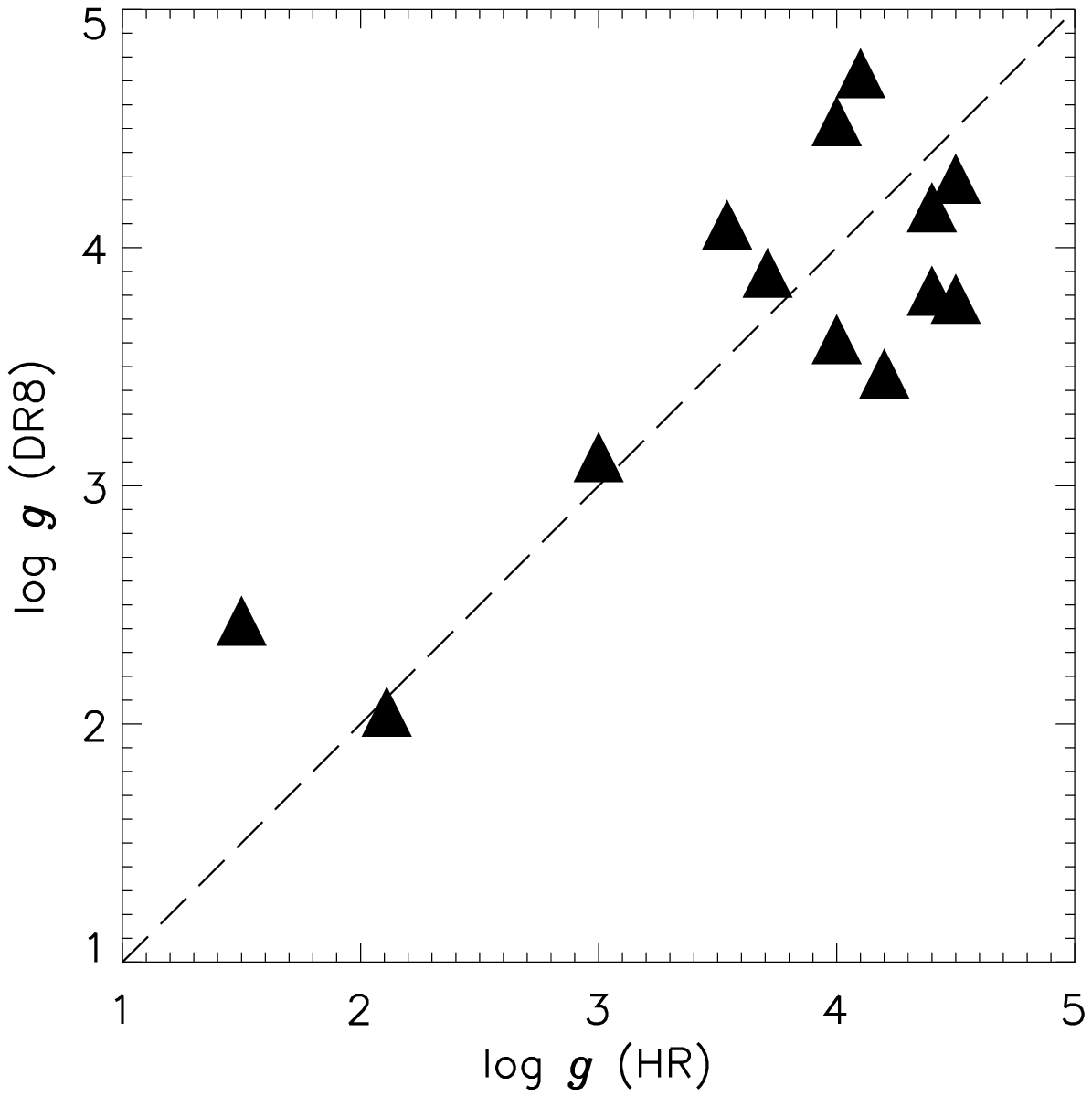}\label{Fig:f1d}}}
   \renewcommand{\captionlabelfont}{\bf}
   \captionsetup{labelsep=space}
   \caption{\small (a) Comparison of $T_1$ and $T_2$ effective temperatures deduced from the (\emph{g} $-$ \emph{z})$_0$  color and those adopted in high-resolution analyses for the 13 metal-poor sample stars. The effective temperatures given in SDSS DR8 are plotted in (b). (c) Comparison of surface gravities \textrm{log}~\emph{g} obtained from the isochrones and those adopted in high-resolution analyses. (d) Same as (b) but for surface gravities \textrm{log}~\emph{g}. }
\end{figure}
\subsubsection{Surface Gravity}
\label{subsect:Detsugra}
\begin{figure}[htbp]
   \centering
   \includegraphics[width=0.7\textwidth]{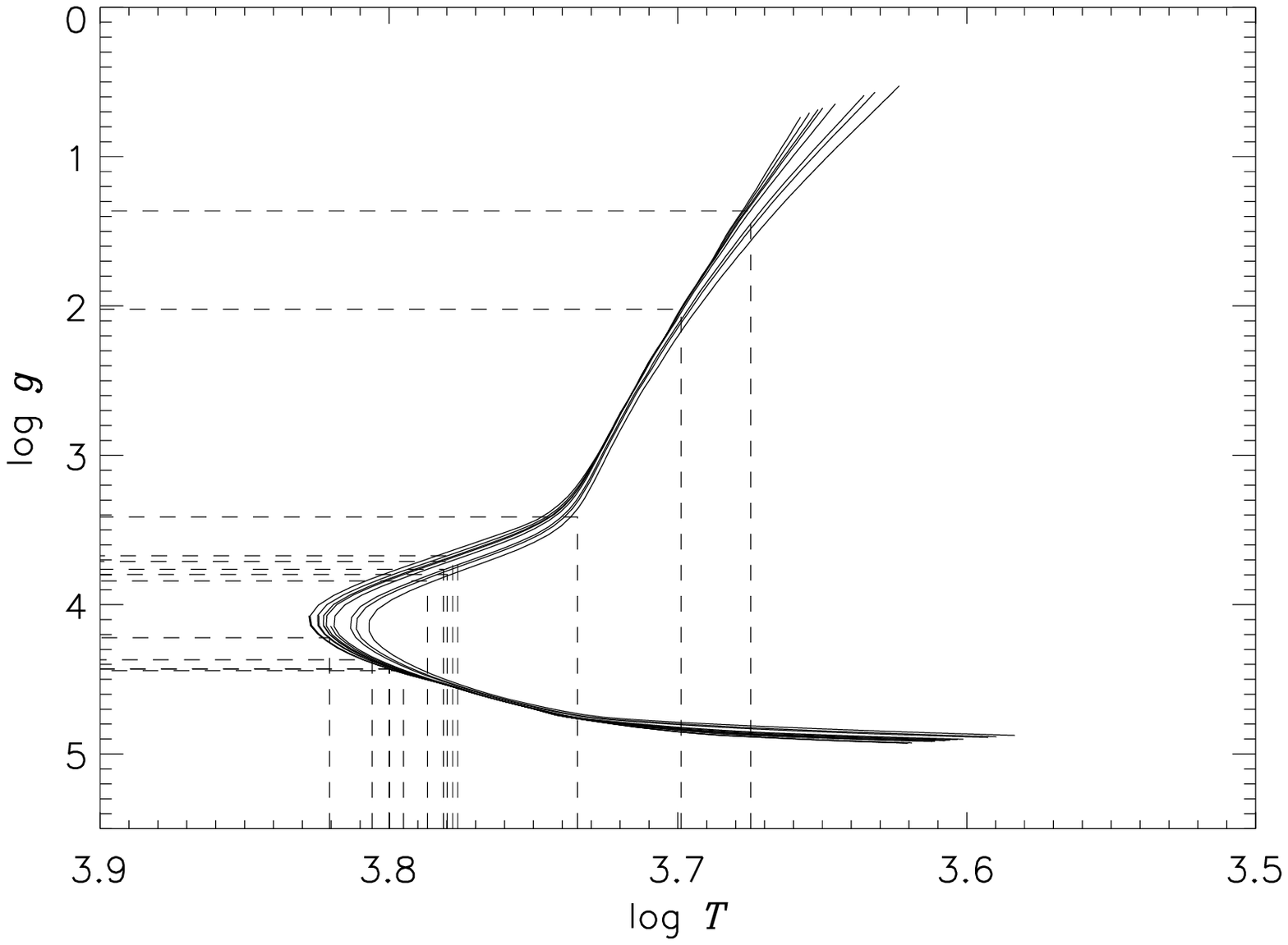}
   \renewcommand{\captionlabelfont}{\bf}
   \captionsetup{labelsep=space}
   \caption{\small The $12$\,Gyr Yonsei-Yale isochrones(\cite{Dema04}) for the 13 metal-poor sample stars, adopting the metallicities given in SDSS DR8 as initial metallicities. The isochrones (solid lines) are in order of increasing metallicity from left ([Fe/H] $= -3.3$) to right ([Fe/H] $= -2.1$). The dashed lines show the value of surface gravity corresponding to the effective temperature estimated for each star. For cooler stars with \textrm{log}~\emph{T}$_{\rm eff}$$ <3.79$, we take the upper parts of the isochrones (above the turn-off points), while for hotter stars with \textrm{log}~\emph{T}$_{\rm eff}$$ >3.79$, we take the lower parts of the isochrones (below the turn-off points).   }
   \label{Fig:f2}
   \end{figure}

We adopt the Yonsei-Yale isochrones
(\cite{Dema04})
for the 13 metal-poor stars. Considering that most metal-poor stars are old stars, we adopt an age of $12$\,Gyr for all of our stars. As the metallicity cannot be easily derived by measuring the equivalent width of faint Fe\,\textsc{\rmnum{1}} lines in the SDSS spectra, we use the metallicity estimates in SDSS DR8 as initial values. The values of surface gravity obtained from the isochrone assuming the effective temperature calculated above are then set as the initial values of stellar surface gravity (see Fig.~\ref{Fig:f2}). Guided by results adopted in the high-resolution spectroscopic analyses, for cooler stars with \textrm{log}~\emph{T}$_{\rm eff}$$ <3.79$, we use the upper parts of the isochrones (above the turn-off points), while for hotter stars with \textrm{log}~\emph{T}$_{\rm eff}$$ >3.79$, we take the lower parts of the isochrones (below the turn-off points). A comparisons of values of surface gravity thus derived with those adopted in the high resolution spectroscopic analyses is given in Fig.\,~\ref{Fig:f1c}.\\
Comparing to values provided in SDSS DR8, our estimates of the surface gravity show much less scatter with those adopted in the high-resolution spectroscopic analyses.

\subsubsection{Microturbulent Velocity}
Given the limited resolution and $S/N$ of the SDSS spectra, we derive the microturbulent velocity by its dependence on effective temperature and surface gravity. Guided by values adopted in the high-resolution spectroscopic analyses, we determine the microturbulence values using Equation (9) in
Edvardsson (\cite{Edva93})
\rm for~$\textrm{log}~\emph{T}_{\rm eff}({\rm K})\geq 3.79$. For $\textrm{log}~\emph{T}_{\rm eff}({\rm K})< 3.79 $, a constant microturbulent velocity $\emph{v}_{\rm turb}=1.5 \rm ~km~s^{-1}$ is adopted.

\subsubsection{Model Atmospheres}
We utilize the local thermodynamical equilibrium (LTE) and line-blanketed Kurucz model atmospheres\footnote{\it http://kurucz.harvard.edu/}. By interpolation in the discrete grids of model atmospheres, we obtain model atmospheres corresponding to the effective temperatures and surface gravities derived above for our sample stars. We employ the program \textsc{abontest8} for abundance analysis. \textsc{abontest8} assumes LTE, and includes effects of natural broadening, thermal broadening, van der Waals damping and microturbulent broadening
(\cite{zz05}).

\subsection{Derivation of Metallicity}
\label{sect:Met}

Using the photometric temperatures and the EW of the line as input, and [Fe/H] from DR8 as initial metallicity, we use the \textsc{abontest8} program to calculate the initial estimate of the $\alpha$-element abundance. The iron abundance is then calculated assuming [$\alpha$/Fe] $= +0.4$. This new iron abundance is then used to recalculate the isochrone, the surface gravity and model atmosphere. The new set of surface gravity and model atmosphere, along with the new iron abundance, are then used as the second set of input parameters to \textsc{abontest8}. The process is iterated, until the derived iron abundance differs from the last input value by less than $0.1$\,dex.\\
There is no significant difference between the metallicities estimated using the two sets of photometric effective temperatures $T_1$ and $T_2$. For the Ca\,\textsc{\rmnum{2}}\,K line, $T_2$ provides results that are in slightly better agreement with the high-resolution spectroscopic values. Accordingly, we have adopted the metallicity values derived from $T_2$.

\subsection{Decision on Final Metallicity Results of the 13 Stars}
\label{sect:Ana}
\begin{figure}[h]
   \centering 
   \mbox{
   \subfigure[]{
   \includegraphics[width=8cm]{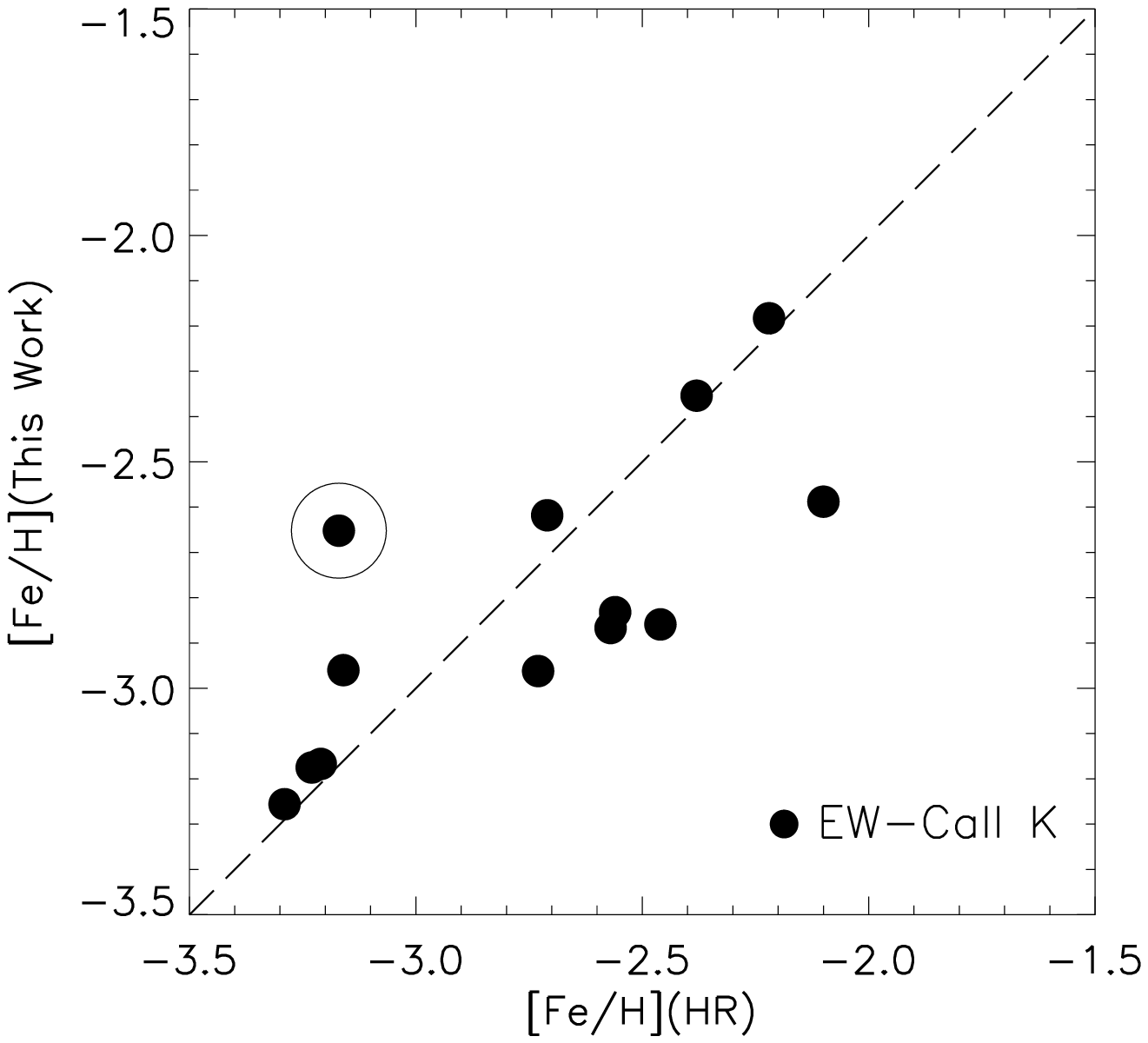} \label{Fig:f3a}}
   \quad
   \subfigure[]{
   \includegraphics[width=8cm]{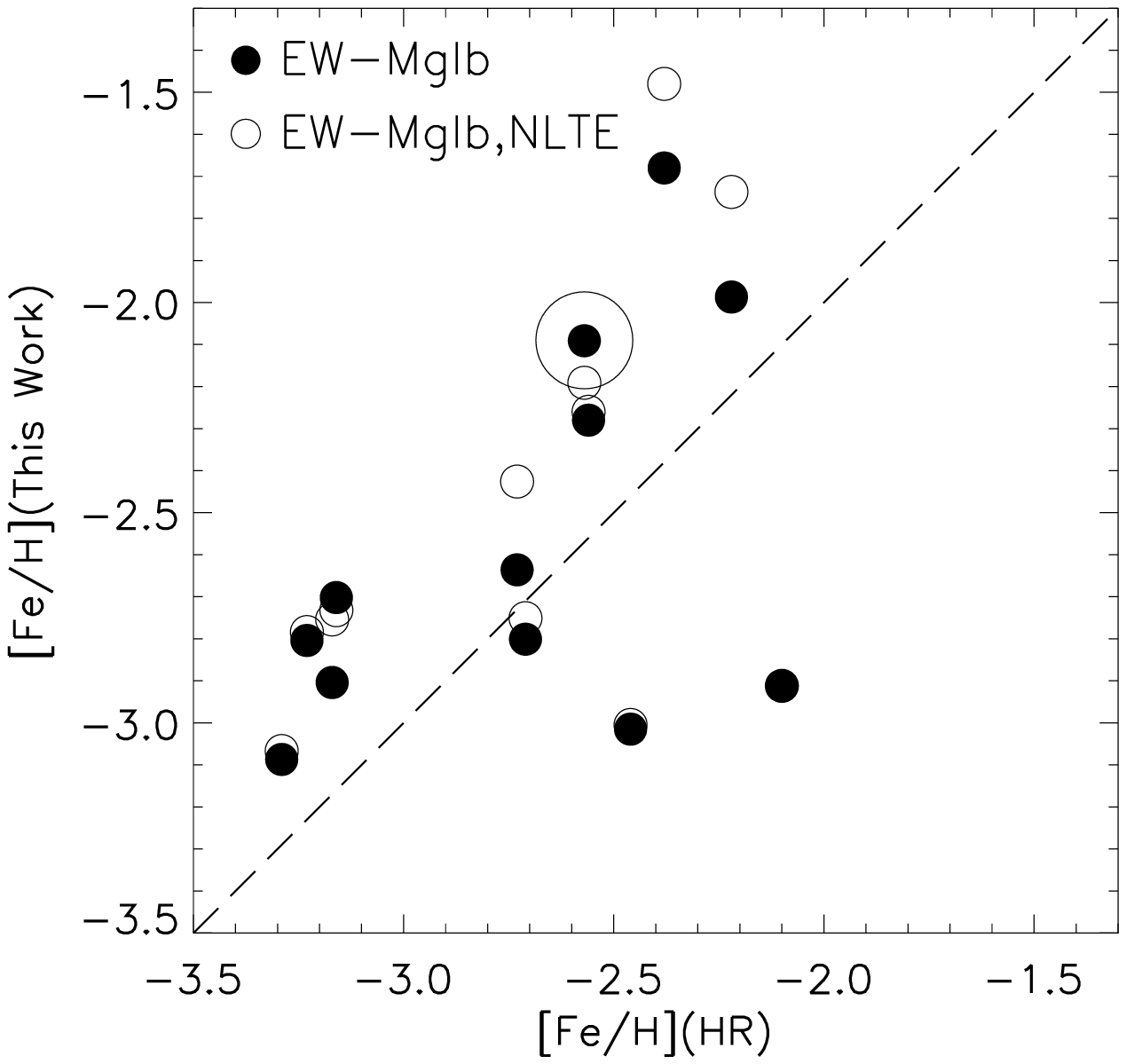} \label{Fig:f3b}}}
   \mbox{
   \subfigure[]{
   \includegraphics[width=8cm]{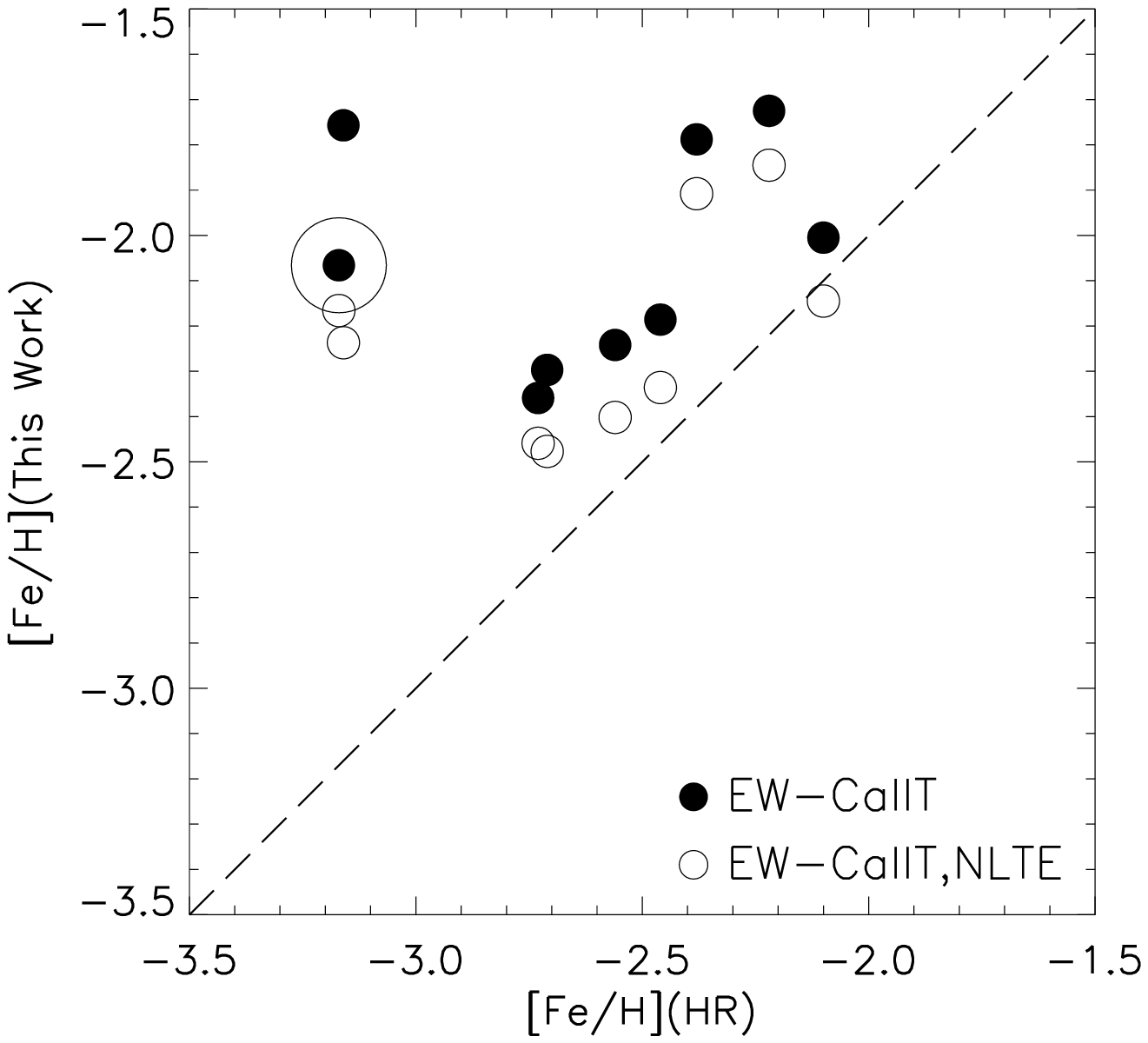} \label{Fig:f3c}}
   \quad
   \subfigure[]{
   \includegraphics[width=8cm]{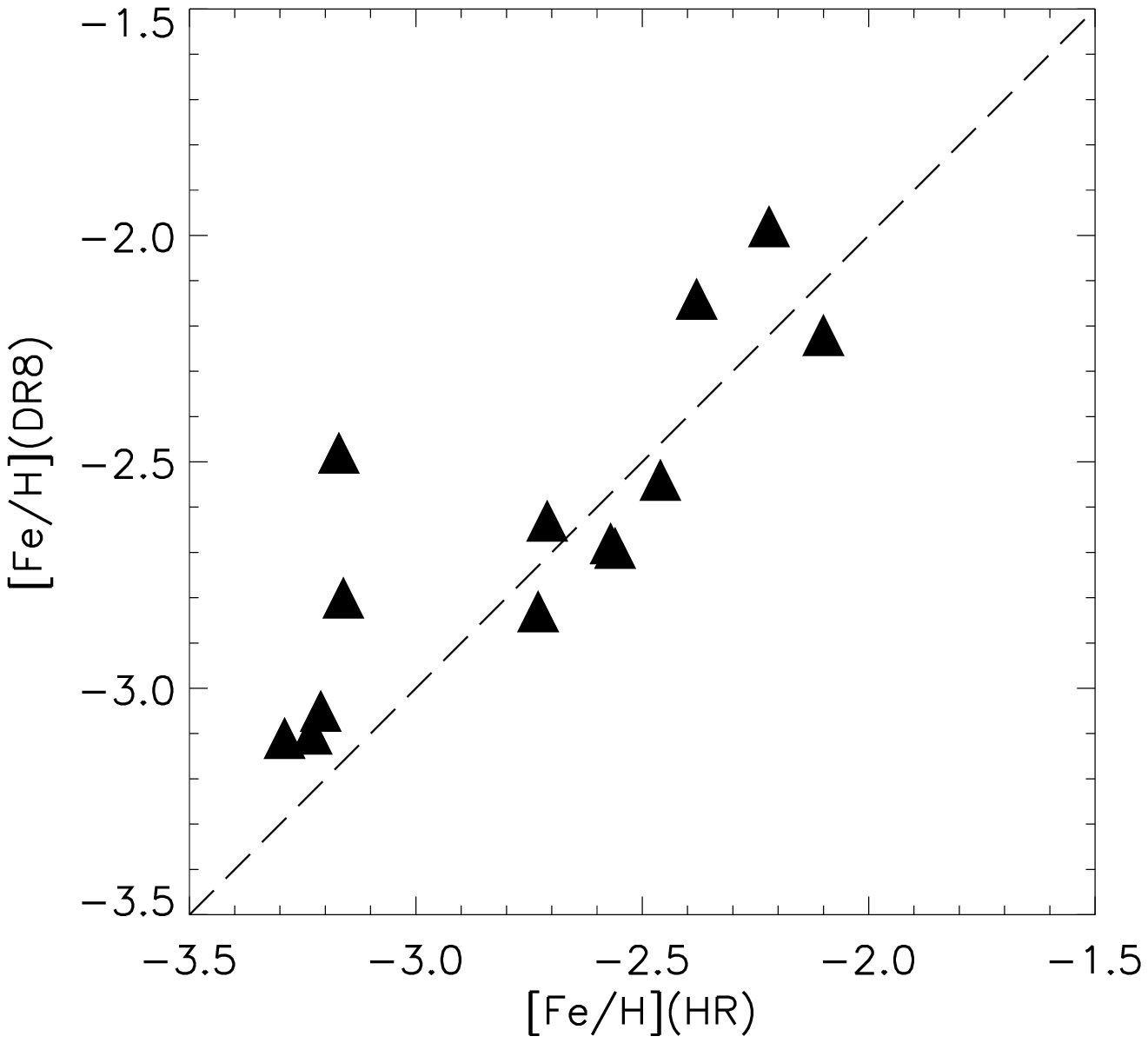} \label{Fig:f3d}}}
   \renewcommand{\captionlabelfont}{\bf}
   \captionsetup{labelsep=space}
   \caption{\small (a), (b), (c) Comparison of our metallicities of the 13 stars derived from the EW of the Ca\,\textsc{\rmnum{2}}\,K line, Mg\,\textsc{\rmnum{1}}\,b lines and Ca\,\textsc{\rmnum{2}}\,triplet lines with the values determined by high-resolution spectra. (d) shows the comparison between SDSS DR8 and high-resolution results as a reference. (b) and (c)  also show our results after NLTE corrections. The big circles indicate stars with large over-abundances of Ca or Mg over Fe.}
   \end{figure}

Our results show that only the metallicities derived from the Ca\,\textsc{\rmnum{2}}\,K line agree well with those determined in high-resolution spectroscopic analyses. The metallicities estimated from the Mg\,\textsc{\rmnum{1}}\,b lines show a large scatter relative to the high-resolution spectroscopic results (see Fig.~\ref{Fig:f3b}). And the metallicities deduced from the Ca\,\textsc{\rmnum{2}}\,triplet lines are much too high (see Fig.~\ref{Fig:f3c}). The Ca\,\textsc{\rmnum{2}}\,K line is the strongest metal absorption line over the optical wavelength range for most stars, while the Mg\,\textsc{\rmnum{1}}\,b and the Ca\,\textsc{\rmnum{2}}\,triplet lines have much weaker line strengths. For the metal-poor stars of interest here, given the low resolution and limited $S/N$ of the SDSS spectra, these metal lines are difficult to measure due to their weakness (the EWs can be as low as $\sim 100$\,m\r{A}), and small errors in fitting the continuum can lead to large errors in the line strength estimates. The effects are particularly profound for weak features. Apart from measurement uncertainties, NLTE effects may also account for parts of the abundance estimate differences, especially for the Ca\,\textsc{\rmnum{2}}\,triplet lines
(Mashonkina et al. ~\cite{Mash07}; \cite{Stark10}). 
To account for this, we have used the methods presented in
Gehren et al. (\cite{Geh06})
and
Mashonkina et al. (\cite{Mash07})
to perform NLTE calculations for the Mg\,\textsc{\rmnum{1}}\,b lines and Ca\,\textsc{\rmnum{2}}\,triplet lines, respectively.\\
Figs.~\ref{Fig:f3a}, \ref{Fig:f3b}, and \ref{Fig:f3c} compare the metallicities derived respectively from the Ca\,\textsc{\rmnum{2}}\,K, Mg\,\textsc{\rmnum{1}}\,b and Ca\,\textsc{\rmnum{2}}\,triplet lines, without and with the NLTE effect corrections. The abundances determined from the Ca\,\textsc{\rmnum{2}}\,triplet lines remain higher than the high-resolution values, even after the NLTE corrections. The discrepancies are probably due to measurement uncertainties. The big circles in the plot indicate the two stars that are very enhanced in [Ca/Fe] or [Mg/Fe]. SDSS J234723 has a [Ca/Fe] ratio of $+1.11$
(\cite{Lai09})
and SDSS1707+58 has [Mg/Fe]$=+1.13$
(\cite{Aoki08}).
Such high [$\alpha$/Fe] values can also contribute to the discrepancies in the estimated [Fe/H]. Table~\ref{Tab:comp-feh} summarizes the averages and standard deviations of the metallicities of our sample derived from each metal line, along with those from SDSS DR7 and DR8. Table~\ref{Tab:comp-feh} shows that the iron abundances deduced from the Ca\,\textsc{\rmnum{2}}\,K line have a higher consistency with the high-resolution spectroscopic results than those derived from the other two sets of metal lines. Therefore, we have adopted the metallicities based on the measured EWs of the Ca\,\textsc{\rmnum{2}}\,K line as the final iron abundances of our sample. The last three columns of Table~\ref{Tab:HR-13}) list the effective temperatures ($T_2$), surface gravities and iron abundances obtained in this work.\\
\begin{table}[h]
\centering
\begin{threeparttable}
\caption[]{Averages and standard deviations of the differences between the [Fe/H] estimated from the different metal lines, those given in DR7 and DR8 and those from the high-resolution spectroscopic analysis of the 13 stars. }\label{Tab:comp-feh}
  \begin{tabular}{c|c|p{1cm}<{\centering}|p{1cm}<{\centering}|p{1cm}<{\centering}|p{1cm}<{\centering}|c|c}
    \hline
 & \multirow{2}{*}{\centering Ca\,\textsc{\rmnum{2}}\,K line} & \multicolumn{2}{c}{Mg\,\textsc{\rmnum{1}}\,b lines} & \multicolumn{2}{|c|}{ Ca\,\textsc{\rmnum{2}}\,triplet lines} & \multirow{2}{*}{\centering SDSS DR7} & \multirow{2}{*}{\centering SDSS DR8}\\
    \cline{3-6}
                           &  & LTE & NLTE & LTE & NLTE & &  \\
    \hline
Average & $0.2$ & $0.4$ & $0.4$ & $0.6$ & $0.5$ & $0.3$ & $0.2$   \\
    \hline
Standard Deviation & $0.3$ & $0.5$ & $0.5$  & $0.8$ & $0.6$ & $0.4$ & $0.3$ \\
    \hline
    \end{tabular}
\end{threeparttable}
\end{table}

\section{Searching for EMP Star Candidates}
\label{sect:Sear}

As shown above, our method that makes use of the Ca\,\textsc{\rmnum{2}}\,K line performs well in returning a reliable and accurate estimate of metallicity, that is, the method is effective in measuring the metallicities of EMP stars using low-resolution spectra. We therefore apply this method to a large sample of metal-poor stars selected from the SDSS archive.

\subsection{Selection of the Sample and Measurement of EW Values}
We select 166 EMP stars (including those with repeated observations) from the SDSS archive which have [Fe/H] $\leq-3.5$, as listed in DR8, and spectra with the average $S/N$ per pixel of more than $20$ over the wavelength range $4000-8000$\,\r{A}. Hot stars, with effective temperatures higher than 7500\,K, are excluded. Using the method described above, we derive their metallicities based on measured line EWs. The results derived from the weak Mg\,\textsc{\rmnum{1}}\,b and Ca\,\textsc{\rmnum{2}}\,triplet lines are not used for the reasons described above.\\
For EMP stars with [Fe/H] $\leq-3.5$, the line features are very weak. Even for the relatively strong Ca\,\textsc{\rmnum{2}}\,K line, it is difficult to identify the feature and fit the continuum using low-resolution spectra. Also, it should be noted that the assumption of [$\alpha$/Fe] $= +0.4$ may lead to overestimated [Fe/H] for stars with strongly enhanced $\alpha$ element. Nevertheless, we have managed to measure the EWs of the Ca\,\textsc{\rmnum{2}}\,K line and determine the metallicities for 45 stars out of the 166 EMP candidates.

\subsection{Results and Discussions}
From the results above, we have identified six metal-poor star candidates with [Fe/H] $\sim -4.0$ using our method. It is noteworthy that one of them, SDSS J102915+172927, was recently confirmed to have a metallicity as low as [Fe/H] $= -4.73$ (based on a 1D model atmosphere analysis) using the Very Large Telescope (VLT) high-resolution spectroscopic observations (\cite{Caff11}). The discovery of this low-mass and most metal-poor star, without the enrichment of C, N and O, sheds light on the theory of formation of the first generation of low-mass stars in the Milky Way. Compared to the VLT result, our iron abundance is overestimated by $\sim 0.6$\,dex. The discrepancy is mainly caused by contamination of the stellar Ca\,\textsc{\rmnum{2}}\,K line by interstellar absorption, which cannot be resolved in the low-resolution SDSS spectra. The other EMP star candidates with [Fe/H] $< -3.8$ in our sample remain to be confirmed. Table~\ref{Tab:ump} lists the stellar parameters of the six EMP star candidates that we have found. \\
\begin{table}[h]
\centering
\begin{threeparttable}
\caption[]{Parameters of the six EMP Star Candidates. }\label{Tab:ump}
\setlength\tabcolsep{2.3pt}
  \begin{tabular}{lccccccc}
    \toprule
STAR & \tabincell{c}{SDSS\\ J102915.14\\+172927.9} & \tabincell{c}{SDSS\\ J144514.69\\+121152.0} & \tabincell{c}{SDSS\\ J100958.70\\+010313.1} & \tabincell{c}{SDSS\\ J144256.37\\-001542.7} & \tabincell{c}{SDSS\\ J230959.55\\+230803.0} & \tabincell{c}{SDSS\\ J161956.33\\+170539.9} \\
    \midrule
RA & $10^{\rm h}29^{\rm m}15.14^{\rm s}$ & $14^{\rm h}45^{\rm m}14.69^{\rm s}$ & $10^{\rm h}9^{\rm m}58.7^{\rm s}$ & $14^{\rm h}42^{\rm m}56.37^{\rm s}$ & $23^{\rm h}9^{\rm m}59.55^{\rm s}$ & $16^{\rm h}19^{\rm m}56.33^{\rm s}$ \\
Dec. & $17\degr29\arcmin27.9\arcsec$ & $12\degr11\arcmin52\arcsec$ & $ 1\degr3\arcmin13.18\arcsec$ & $ -15\arcmin42.74\arcsec$ & $ 23\degr8\arcmin3.06\arcsec$ & $ 17\degr5\arcmin39.93\arcsec$ \\
\emph{T}$_{\rm eff}$ (DR8) & $5765$ & $5150$ & $4462$ & $6129$ & $6294$ & $6314$ \\
\textrm{log}~\emph{g} \rm (DR8) & $3.1$ & $1.9$ & $1.6$  & $3.8$ & $3.3$ & $3.6$ \\
\rm [Fe/H] (DR8) & $-3.8$ & $-3.5$ & $-3.7$ & $-3.7$ & $-3.6$ & $-3.5$ \\
\emph{T}$_{\rm eff}$ (This work) & $5775$ & $4464$ & $4435$ & $6110$ & $5908$ & $6138$ \\
\textrm{log}~\emph{g} \rm (This work) & $3.4$  & $0.5$ & $0.5$ & $3.6$ & $3.5$ & $3.7$ \\
\rm [Fe/H] (This work) & $-4.1$ & $-4.0$ & $-3.9$ & $-3.8$ & $-3.9$ & $-3.9$ \\
    \bottomrule
    \end{tabular}
\end{threeparttable}
\end{table}

\section{Summary}
\label{sect:Con}
Using the EW of the Ca\,\textsc{\rmnum{2}}\,K line, effective temperature deduced from the (\emph{g} $-$ \emph{z})$_0$ color and surface gravity obtained from the isochrone, we can derive a reliable estimate of metallicity for EMP stars with the SDSS spectra.\\
We find the metallicities derived from the Ca\,\textsc{\rmnum{2}}\,K line have good agreement with the high-resolution spectroscopic results. From a sample of 166 potential EMP stars selected from SDSS DR8, we identify six metal-poor star candidates with [Fe/H] $\sim -4.0$, and one of them, SDSS J102915+172927, has recently been confirmed to be ultra metal-poor ([Fe/H] $< -4.0$) by the VLT observations
(\cite{Caff11}), showing our method is effective even at the lowest known metallicities.\\
Further high-resolution spectroscopic analyses for the other stars with estimated metallicities [Fe/H] $\sim -4.0$ in our sample is of considerable interest. With a rich source of low-resolution spectra provided by the ongoing SDSS and the forthcoming LAMOST campaigns, our method has a great potential to discover more EMP stars in the near future.

\begin{acknowledgements}
This work was supported by the National Natural Science Foundation of China under Grant No.11073003 and 11078006. \\
Z.H.W. thanks Prof. Mashonkina for providing calcium NLTE model. Z.H.W. also thanks Prof. Shi Jianrong for helping us with model calculations.\\
Funding for the SDSS and SDSS-\Rmnum{2} has been provided by the Alfred P. Sloan Foundation, the Participating Institutions, the National Science Foundation, the U.S. Department of Energy, the National Aeronautics and Space Administration, the Japanese Monbukagakusho, the Max Planck Society, and the Higher Education Funding Council for England. The SDSS Web Site is \it http://www.sdss.org/. \rm The SDSS is managed by the Astrophysical Research Consortium for the Participating Institutions. The participating institutions are the American Museum of Natural History, Astrophysical Institute Potsdam, University of Basel, University of Cambridge, Case Western Reserve University, University of Chicago, Drexel University, Fermilab, the Institute for Advanced Study, the Japan Participation Group, Johns Hopkins University, the Joint Institute for Nuclear Astrophysics, the Kavli Institute for Particle Astrophysics and Cosmology, the Korean Scientist Group, the Chinese Academy of Sciences (LAMOST), Los Alamos National Laboratory, the Max-Planck-Institute for Astronomy (MPIA), the Max-Planck-Institute for Astrophysics (MPA), New Mexico State University, Ohio State University, University of Pittsburgh, University of Portsmouth, Princeton University, the United States Naval Observatory, and the University of Washington.\\
\end{acknowledgements}

\label{lastpage}
\end{document}